\documentclass[sigconf]{acmart}

\usepackage{booktabs}
\usepackage{multirow}
\usepackage{graphicx}
\usepackage{color}
\usepackage{bm}
\usepackage{amsmath}
\usepackage{dsfont}
\usepackage{amssymb}
\usepackage{lipsum}
\usepackage{setspace}

\AtBeginDocument{%
  \providecommand\BibTeX{{%
    \normalfont B\kern-0.5em{\scshape i\kern-0.25em b}\kern-0.8em\TeX}}}

\copyrightyear{2020} 
\acmYear{2020} 
\setcopyright{acmcopyright}\acmConference[MM '20]{Proceedings of the 28th ACM International Conference on Multimedia}{October 12--16, 2020}{Seattle, WA, USA}
\acmBooktitle{Proceedings of the 28th ACM International Conference on Multimedia (MM '20), October 12--16, 2020, Seattle, WA, USA}
\acmPrice{15.00}
\acmDOI{10.1145/3394171.3413869}
\acmISBN{978-1-4503-7988-5/20/10}

\begin{document}
\fancyhead{}
\title{Look, Listen, and Attend: Co-Attention Network for Self-Supervised Audio-Visual Representation Learning}

\author{Ying Cheng$^{1}$, \ \ Ruize Wang$^{1}$, \ \ Zhihao Pan$^{2}$, \ \ Rui Feng$^{1,2,*}$, \ \ Yuejie Zhang$^{2,*}$}

\affiliation{%
\institution{$^{1}$ Academy for Engineering and Technology, Fudan University, China}
}
\affiliation{\institution{$^{2}$School of Computer Science, Shanghai Key Laboratory of Intelligent Information Processing, Fudan University, China}}

\affiliation{\institution{\{chengy18, rzwang18, zhpan18, fengrui, yjzhang\}@fudan.edu.cn}}

\renewcommand{\shortauthors}{Cheng and Wang, et al.}

\begin{abstract}
When watching videos, the occurrence of a visual event is often accompanied by an audio event, e.g., the voice of lip motion, the music of playing instruments. There is an underlying correlation between audio and visual events, which can be utilized as free supervised information to train a neural network by solving the pretext task of audio-visual synchronization. In this paper, we propose a novel self-supervised framework with co-attention mechanism to learn generic cross-modal representations from unlabelled videos in the wild, and further benefit downstream tasks. Specifically, we explore three different co-attention modules to focus on discriminative visual regions correlated to the sounds and introduce the interactions between them. Experiments show that our model achieves state-of-the-art performance on the pretext task while having fewer parameters compared with existing methods. 
To further evaluate the generalizability and transferability of our approach, we apply the pre-trained model on two downstream tasks, i.e., sound source localization and action recognition. Extensive experiments demonstrate that our model provides competitive results with other self-supervised methods, and also indicate that our approach can tackle the challenging scenes which contain multiple sound sources.
\end{abstract}




\maketitle

\newcommand\blfootnote[1]{%
\begingroup
\renewcommand\thefootnote{}\footnote{#1}%
\addtocounter{footnote}{-1}%
\endgroup
}

\blfootnote{$^{*}$ indicates corresponding authors.}

\begin{small}

\setlength{\parskip}{-2em} 
\setlength{\parindent}{0pt}
\begin{spacing}{1}
\textbf{ACM Reference Format:} \\
Ying Cheng, Ruize Wang, Zhihao Pan, Rui Feng, Yuejie Zhang. 2020. Look, Listen, and Attend: Co-Attention Network for Self-Supervised Audio-Visual Representation Learning. In \emph{Proceedings of the 28th ACM International Conference on Multimedia (MM '20), October 12--16, 2020, Seattle, WA, USA.} ACM, New York, NY, USA, 9 pages. https://doi.org/10.1145/3394171.3413869
\end{spacing}
\end{small}

\section{Introduction}
In many tasks of video analysis, applying supervised learning to train a neural network is very expensive. Annotators need to watch entire videos and manually add labels to each video (even each frame in videos). Only a small amount of video data can be used for training via this learning method. 
However, vast numbers of images/videos are uploaded to social websites every day. It is such a waste if we cannot fully exploit them. 

Self-supervised learning is a framework which aims to learn generic  representations without human-labeling.
The supervised signals come from the data itself, e.g., audio-visual co-occurrences. Audio and visual events tend to occur together in videos. For example, when a person is chopping wood, we can hear the clash between the axe and the wood. The audio-visual co-occurrences provide free supervised signals, which can be utilized for co-training a joint neural network and learning cross-modal representations. Compared with the fully-supervised methods, self-supervised learning provides the opportunity for exploiting large-scale unlabelled data. 

\begin{figure} [!t]
\centering
  \includegraphics[width=0.99\linewidth]{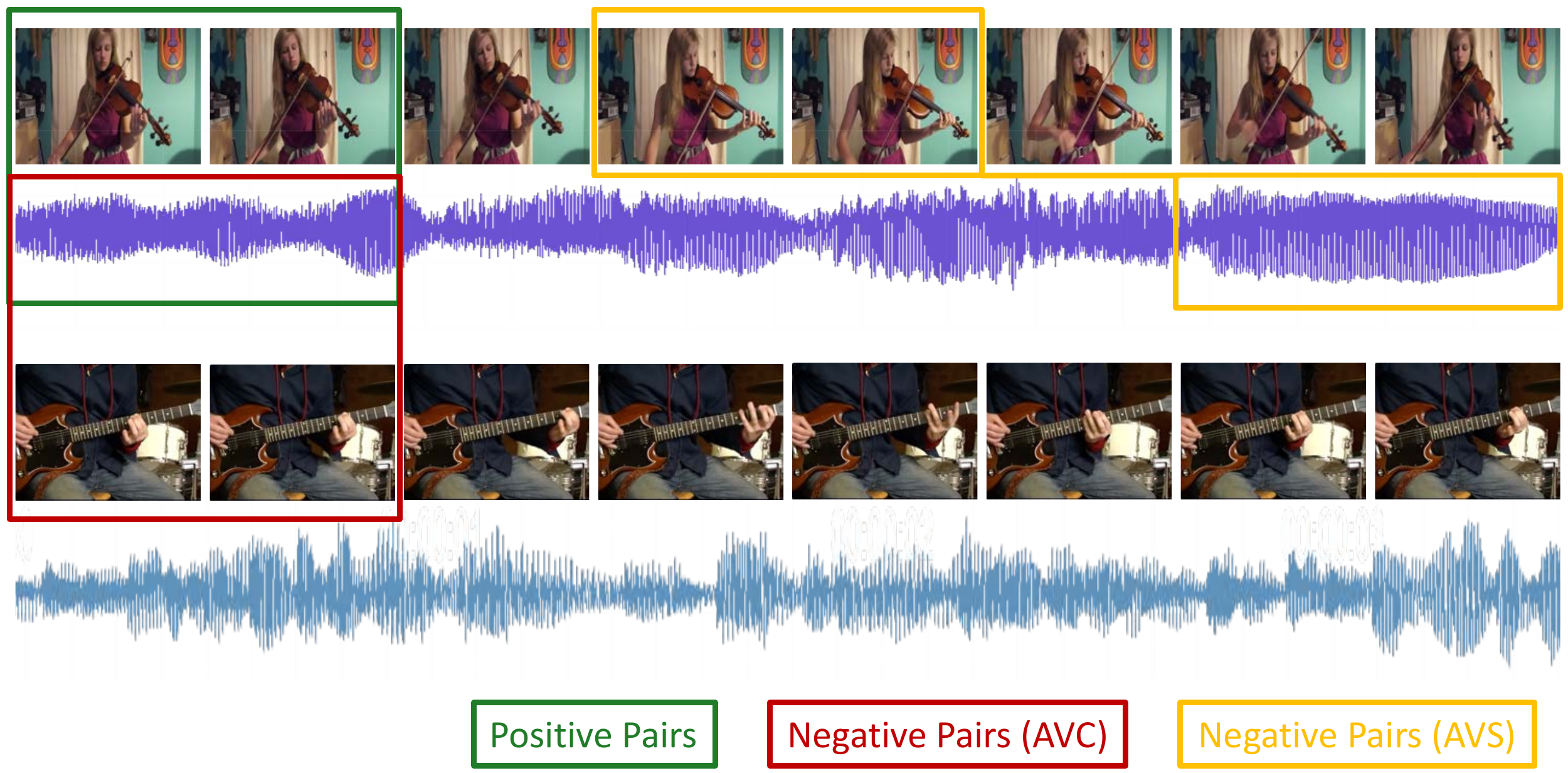}
  \caption{An illustration of different audio-visual pretext tasks, i.e., Audio-Visual Correspondence (AVC) and Audio-Visual Synchronization (AVS). The examples in the green box are positive pairs in the pretext tasks (both AVC and AVS), which are extracted from the same time of one video. The examples in the red box are negative pairs in the AVC task, which are extracted from different videos. The examples in the yellow box are negative pairs in the AVS task, which are extracted from different slices of one video.
}
  \label{fig:intro}
\end{figure}

Recently, some efforts have been made for learning audio and visual semantic representations in a self-supervised way. These methods (also called pretext tasks) can be divided into two categories, i.e., Audio-Visual Correspondence (AVC) and Audio-Visual Synchronization (AVS). The VGG group at Oxford University \cite{arandjelovic2017look,arandjelovic2018objects} proposed the AVC task, i.e., predicting whether visual frames and audio clips are sampled from the same video. As illustrated in Figure \ref{fig:intro}, 
the main difference between the AVC and AVS task is the negative pairs.
The objective of the AVS task is to detect the misalignments between audio and visual streams. Most of previous researchers \cite{chung2016out,chung2019perfect,halperin2019dynamic,nagrani2020disentangled} focused on speech tasks, and thus were limited to the domain of face scenes. \citet{owens2018audio,korbar2018cooperative} proposed to use more general data to learn semantic representations, which could be leveraged in practical applications. However, previous approaches overlook the information exchange between modalities, and they are pervasively limited by the heterogeneous complexity of audio-visual scenes, i.e., multiple sound sources. 

In this paper, we focus on how to learn generic cross-modal representations in the complex scenes efficiently. We note that there is an interesting phenomenon, some nearsighted people are difficult to hear clearly when taking off glasses. This can be explained by the McGurk effect \cite{mcgurk1976hearing}, which demonstrates an interaction between hearing and vision in humans. Inspired by such a specific phenomenon, we propose a self-supervised framework with co-attention mechanism to provide information exchange between audio and visual streams. To jointly attend to distinct sound sources and their visual regions, we also apply the multi-head based structure \cite{vaswani2017attention} to cross-modal attention. After training on the pretext task, we employ the pre-trained model as a base for two audio-visual downstream tasks, i.e., sound source localization and action recognition. Extensive experiments indicate that our model achieves superior performance and can tackle complicated scenes, while having fewer parameters compared with previous approaches.

The contributions of this paper can be summarized as follows:
\begin{itemize}
\item We propose to introduce interactions between audio and visual steams for self-supervised learning, learn cross-modal representations, and eventually benefit downstream tasks.
\item We propose a co-attention framework to exploit the co-occurrences between audio and visual events, in which the multi-head based structure can effectively associate the discriminative visual regions with the sound sources.
\item Extensive experiments indicate that our method has good generalizability and can tackle the challenging scenes which contain multiple sound sources.
\end{itemize}

\section{Related Work}
We first concisely review the works of representation learning for audio and visual modalities, and then discuss the relevant progress in the area of audio-visual applications.
\subsection{Audio-Visual Representation Learning}
In recent years, there have been some works that focus on audio-visual representation learning. These approaches can be divided into three classes according to the source of supervision signals, i.e., vision, audio, and both of them. 
The early works \cite{aytar2016soundnet,owens2016ambient,aytar2017see} argued that the audio and visual representations of the same instance had similar semantic information. Thus they investigated how to transfer supervision between different modalities using unlabeled videos as a bridge. Such "teacher-student" training procedure leverages the discriminative knowledge from a well-trained model of one modality to supervise another different modality. \citet{aytar2016soundnet} used visual information as supervision for acoustic scene/object classification, while \citet{owens2016ambient} exploited ambient sound to supervise the learning of visual representations. Although the aforementioned approaches have shown promising cross-modal transfer learning ability, they still rely on the knowledge from established models with a large amount of training data.

Recently, some researchers are interested in whether the audio and visual modalities can supervise each other. \citet{arandjelovic2017look} proposed to learn audio and visual semantic representations by predicting whether the static image and audio clip correspond to each other. They further developed the Audio-Visual Embedding Network (AVE-Net) \cite{arandjelovic2018objects} to facilitate cross-modal retrieval and sound source localization. \citet{hu2019deep} and \citet{alwassel2019self} introduced multimodal clustering to disentangle each modality into distinct components and performed efficient correspondence learning between the components. \citet{owens2018audio} transformed the pretext task from audio-visual correspondence to temporal synchronization and enforced the models to learn spatio-temporal features. \citet{korbar2018cooperative} proposed to combine these two pretext tasks by using the strategy of curriculum learning. However, these approaches neglect the information exchange between modalities. In contrast, our framework allows the communications of audio and visual information through co-attention modules, and finds the underlying correlations between them.

\begin{figure*} [!ht]
\centering
  \includegraphics[width=1\textwidth]{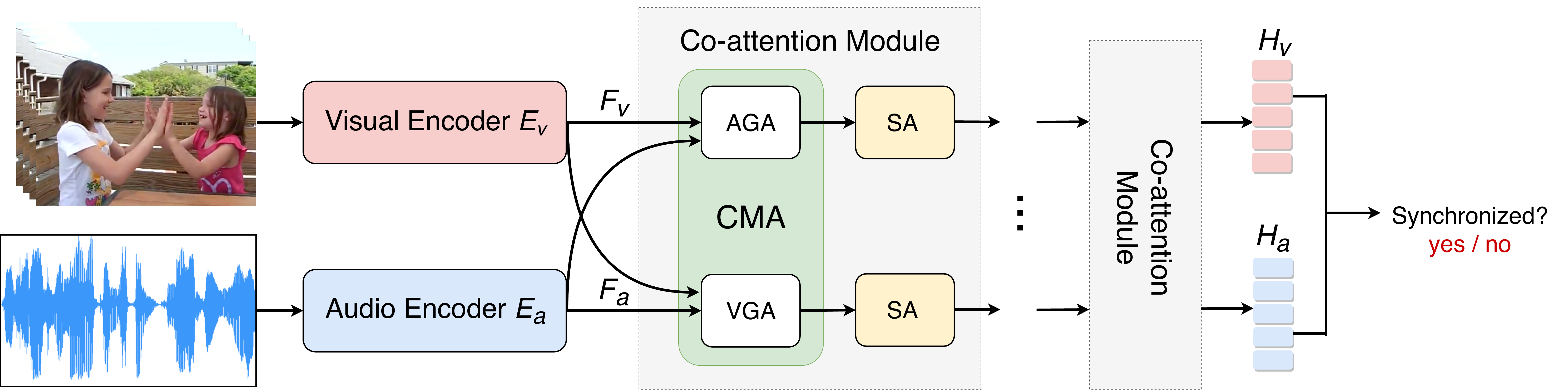}
  \caption{An overview of our co-attention model for the Audio-Visual Synchronization (AVS) task. The co-attention module consists of the CMA (cross-modal attention) block followed by the SA (self-attention) block, in which the CMA block is composed of AGA (audio-guided attention) and VGA (visual-guided attention). The ellipsis denotes that the co-attention modules can be cascaded in depth.
}
  \label{fig:co-attention}
\end{figure*}

\subsection{Audio-Visual Applications}
\paragraph{Sound Source Localization}
The task of sound source localization is to localize visual regions correlated to the sounds.
In the computational approaches, multivariate Gaussian process model \cite{hershey2000audio}, non-parametric methods \cite{fisher2001learning}, Canonical Correlations Analysis (CCA) \cite{kidron2005pixels,izadinia2012multimodal}, and keypoints/probabilistic formalism \cite{barzelay2007harmony} are used to associate the visual regions with the sounds. Besides, some acoustic hardware based approaches \cite{van2004optimum,zunino2015seeing} use specific devices (e.g., microphone arrays) to capture phase differences of sound arrival, utilizing the physical relationships to localize sound sources. With the development of deep learning, more and more researchers have studied this task. Some methods \cite{zhao2018sound,arandjelovic2018objects} focused on the domain of musical instruments localization. On the other hand, several methods \cite{arandjelovic2017look, owens2018audio} explored the sound source localization in more generic scenes. However, such methods are difficult to recognize distinct sources when the video contains mixed multiple sounds. 
Our network with multi-head based structure can localize each sound source in the complicated scenes without any supervision.

\paragraph{Action Recognition}
Action Recognition is a very attractive topic due to its various real-world applications, including visual surveillance \cite{hu2007semantic,singh2010muhavi,ji20123d}, human-computer interaction \cite{ryoo2015robot,koppula2015anticipating}, video retrieval \cite{ciptadi2014movement,ramezani2016review}, etc. Researchers have proposed various approaches to address this problem, such as C3D \cite{tran2015learning}, I3D \cite{carreira2017quo}, R(2+1)D \cite{tran2018closer}, S3D \cite{xie2018rethinking}. However, such models are trained in a fully-supervised way, and normally require a great deal of annotated training data. Although the datasets of Sports-1M \cite{KarpathyCVPR14} and Youtube-8M \cite{abu2016youtube} provide millions of human activity videos, their annotations are generated by the involved main topics automatically, and thus might not be accurate. Recently, some self-supervised methods \cite{korbar2018cooperative,owens2018audio,alwassel2019self} have been proposed to exploit the sufficiency of unlabeled videos and boost the generalizability of models. Compared with these works, our method introduces the interactions between modalities and provides competitive results with fewer parameters of the model. 

\section{Audio-Visual Synchronization}
The audio and visual events tend to occur together at the same time, and the videos provide natural alignments between them. Therefore, it is achievable to determine whether the extracted semantic contents are corresponding by detecting the synchronization of streams in videos, and result in generic audio and visual representations without any manual annotations.

In this paper, the pretext task of cross-modal self-supervised learning is Audio-Visual Synchronization (AVS), which can be set up as a binary classification problem. Taking a set of video clips as inputs, audio and visual streams are synchronized in positive samples while unsynchronized in negative samples, and our models are trained to distinguish between these samples. Specifically, given a training dataset consisting of $N$ video clips $X=\left\{\left(a_{1}, v_{1}, y_{1}\right), \cdots,\left(a_{n}, v_{n}, y_{n}\right)\right\}$, where $a_{n}$ and $v_{n}$ denote the $n$-th audio sample and the $n$-th visual clip, respectively. The label $y_{n} \in\{0,1\}$ indicates whether the audio and visual inputs are synchronized, which is obtained from the video clips directly without human annotations. If $y_{n}=1$, $a_{n}$ and $v_{n}$ are sampled from the same time of the video. If $y_{n}=0$, $a_{n}$ and $v_{n}$ are sampled from different slices of the video.

We consider that this task requires the model to associate discriminative visual regions with the sound sources. To do this, we propose a neural network with co-attention mechanism, as illustrated in Figure \ref{fig:co-attention}. Let $E_{a}$ and $E_{v}$ be the audio and visual encoders, respectively. Thus, we can obtain the set of audio features $F_{a}=\left\{f_{a}^{(n)}=E_{a}\left(a_{n}\right)\right\} \in \mathbb{R}^{d_{a} \times N}$ and visual features $F_{v}=\left\{f_{v}^{(n)}=E_{v}\left(v_{n}\right)\right\} \in \mathbb{R}^{d_{v} \times N}$, where $d_{a}$ and $d_{v}$ denote the dimension of the audio and visual feature, respectively. To introduce the information interactions between audio and visual streams, we explore three different co-attention modules consisting of transformer \cite{vaswani2017attention} blocks, which can be cascaded in depth. We will describe the details of co-attention modules in Section 3.1. Given a set of audio features $F_{a}$ and visual features $F_{v}$, the final output representations of co-attention modules are $H_{a}$ and $H_{v}$, which are flattened, concatenated, and passed to the fully-connected fusion layers to predict the synchronization probability. 

\subsection{Network Architecture}
The architecture of our network is composed of three main parts: audio and visual encoders which are used for extracting audio and visual features, respectively, and co-attention modules that provide interactions between modalities. 

\paragraph{Audio encoder.}
For the audio stream, we follow the approach in \cite{owens2018audio}. The input waveforms are ingested to a sequence of 1D convolutional filters, where $g_o$, $g_d$ and $g_s$ denote the filter output dimension, filter size, and filter stride, respectively. The waveform is 1D ($N$ samples by 2 stereo channels) but reshaped to [$N$, 1, 1, 2], and thus the convolution is implemented as 3D convolution. Specifically, the audio encoder consists of one 3D convolutional filter ($g_o$=64, $g_d$=[65*1*1], $g_s$=4), one 3D average pooling ($g_o$=64, $g_d$=[4*1*1], $g_s$=[4*1*1]), two 3D convolutional filters ($g_o$=128, $g_d$=[15*1*1], $g_s$=4), two 3D convolutional filters ($g_o$=128, $g_d$=[15*1*1], $g_s$=4), two 3D convolutional filters ($g_o$=256, $g_d$=[15*1*1], $g_s$=4), one 3D average pooling ($g_o$=128, $g_d$=[3*1*1], $g_s$=[3*1*1]), and one 3D convolutional filter ($g_o$=128, $g_d$=[3*1*1], $g_s$=1). A fully connected layer is applied to obtain the 512D audio features $F_a$.

\paragraph{Visual encoder.}
In order to capture the motion information, the visual encoder consists of a small number of 3D convolutional filters, where $g_o$, $g_d$ and $g_s$ denote the filter output dimension, filter size, and filter stride, respectively. First, we reshape the inputs into a size of $t$ $\times$ 224 $\times$ 224 $\times$ 3, where $t$ denotes the number of frames. We then feed the input features into one 3D convolutional filter ($g_o$=64, $g_d$=[5*7*7], $g_s$=1), perform 3D average pooling ($g_o$=64, $g_d$=[1*3*3], $g_s$=[1*2*2]), and go through four 3D convolutional filters ($g_o$=64, $g_d$=[3*3*3], $g_s$=2). At last, we average all frame features and apply a fully connected layer to obtain the 512D video features $F_v$.

\begin{figure*} [!ht]
\centering
  \includegraphics[width=0.9\textwidth]{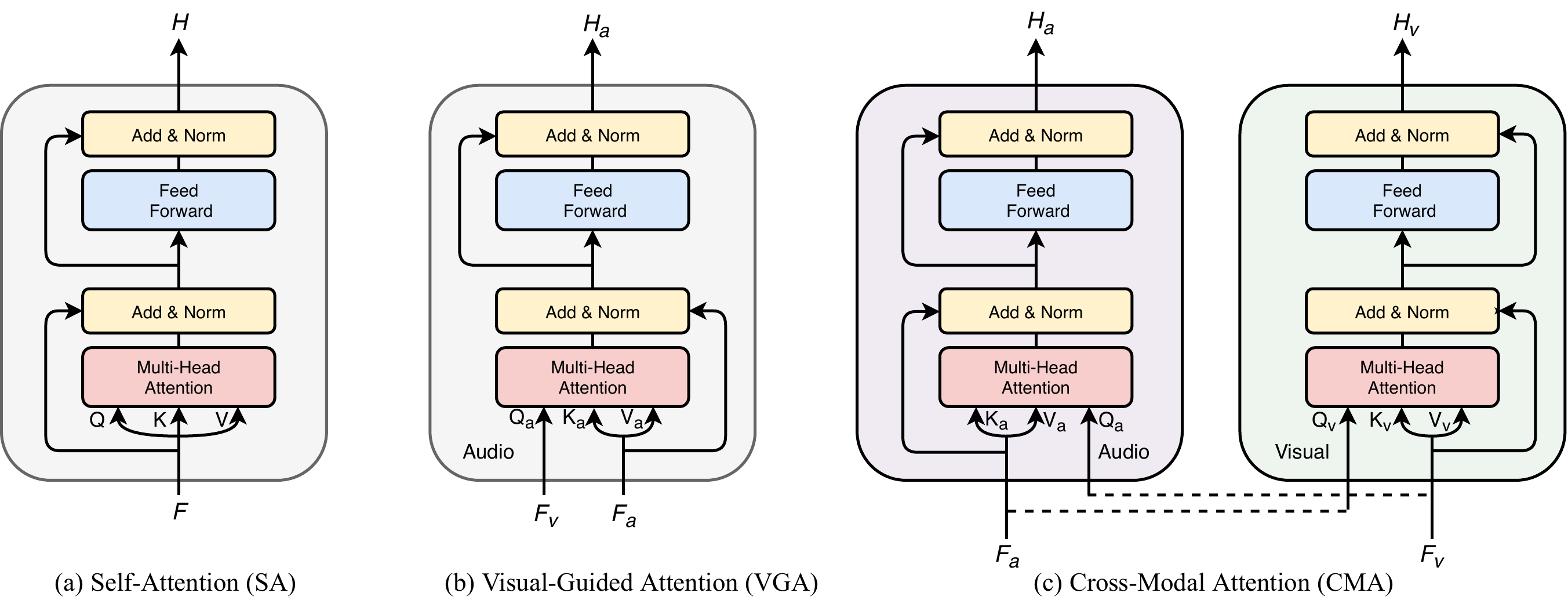}
  \caption{Three basic blocks of co-attention modules. (a) Self-Attention (SA), in which the vectors of queries (Q), keys (K) and values (V) are same as the input features; (b) Visual-Guided Attention (VGA), in which the inputs of $Q_{a}$ are modified to the visual features $F_{v}$; (c) Cross-Modal Attention (CMA), in which the queries of each modality are modified to the features of the other modality. These blocks introduce the interactions between modalities, and the multi-head attention structure enables the model to focus on discriminative concrete components.
}
  \label{fig:transformer}
\end{figure*}

\paragraph{Co-attention module.}
Before presenting the architecture of co-attention modules, we first introduce the Self-Attention (SA) encoder block in the transformer. The inputs of the SA block are the vectors of queries ($Q$), keys ($K$), and values ($V$). The output attended features are obtained by weighted summation over the values $V$, as formulated in Eq. (1). The weights on the values are computed as the dot-product similarity between the queries $Q$ and keys $K$, divide each by $\sqrt{d}$, where $d$ denotes the dimension of $Q$, $K$ and $V$.

\begin{equation}
\operatorname { Attention }(Q, K, V)=\operatorname{softmax}\left(\frac{Q K^{T}}{\sqrt{d}}\right) V
\end{equation}

To further focus on discriminative concrete components, multi-head attention structure is adopted in the blocks, which consists of $m$ paralleled "heads". Regarding each independent attention function as one head, the output features of each head are concatenated and then projected into to one value, yielding the final output as:

\begin{equation}
\begin{aligned}
\operatorname{ MultiHead }(Q, K, V) &\left.=\text { Concat (head }_{1}, \ldots, \text { head }_{\mathrm{m}}\right) W^{O} \\
\text { where head}_{\mathrm{i}} &=\operatorname{ Attention }\left(Q W_{i}^{Q}, K W_{i}^{K}, V W_{i}^{V}\right)
\end{aligned}
\end{equation}
where $W_{i}^{Q}, W_{i}^{K}, W_{i}^{V} \in \mathbb{R}^{d \times d_{m}}$ and $W^{O} \in \mathbb{R}^{m \times d_{m} \times d}$ are the learned projection matrices. $d_{m}$ is the dimension of the output from each head. To prevent the dimension of output feature becoming too large, $d_{m}$ is usually set to $d / m$.

As shown in Figure 3(a), the SA block consists of a multi-head attention structure followed by a position-wise fully connected feed-forward network. Residual connection \cite{he2016deep} and layer normalization \cite{ba2016layer} are also applied to each of the two sub-layers. Specifically, in the SA block, the vectors of $Q$, $K$ and $V$ are the same, which are equal to the intermediate representations $F$. Taking the set of representations $F=\left(f_{1}, \cdots, f_{n}\right) \in \mathbb{R}^{d \times N}$ as inputs, we can obtain the attended output features $H=\left(h_{1}, \cdots, h_{n}\right) \in \mathbb{R}^{d \times N}$. 

Inspired by the McGurk effect, we first propose a Visual-Guided Attention (VGA) transformer block, in which the intermediate visual features $F_{v}$ guide the attention learning for audio stream. As depicted in Figure 3(b), the query vectors of audio steam $Q_{a}$ passed to the multi-head attention are modified to the visual features $F_{v}$ rather than the extracted audio features $F_{a}$. Thus, the attention maps of the VGA block tend to focus on the values in the audio stream related to visual information. Similar to the VGA block, we also design an Audio-Guided Attention(AGA) block, in which the audio features $F_{a}$ guide the attention learning for visual stream and eventually obtain the attended visual features $H_{v}$. In addition to the VGA and AGA blocks, we further explore a Cross-Modal Attention (CMA) block for the AVS task. As depicted in Figure 3(c), the queries of each modality are modified to the intermediate features of the other modality, which introduce cross-modal interactions between audio and visual streams. These interactions are further exploited to obtain the final attended audio features $H_{a}$ and visual features $H_{v}$. It should be noted that such three attention blocks are all followed by the SA blocks to model the intra-modal interactions in audio and visual streams, and these attention blocks can be cascaded in depth to gradually refine the attended cross-modal features. 

\begin{figure*} [!ht]
\centering
  \includegraphics[width=0.9\textwidth]{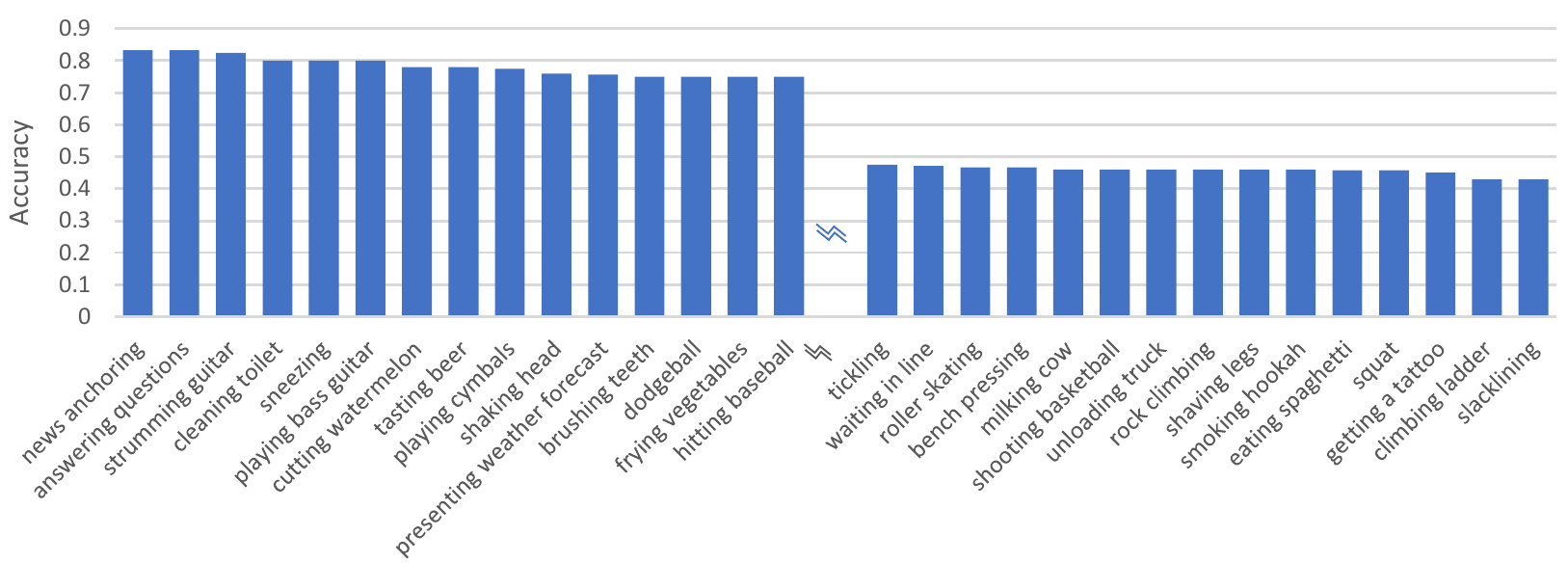}
  \caption{The transfer experimental results on the Audio-Visual Synchronization (AVS) task of action categories in the Kinetics dataset. The categories are sorted from high to low according to the accuracy.
}
  \label{fig:shift_test}
\end{figure*}

\subsection{Implementation Details}
\paragraph{Datasets.}
Audioset \cite{45857} contains 2,084,320 manually-annotated 10-sec. segments of 632 audio event classes from YouTube. For efficiency purposes, we train our neural network on a subset of approximately 240K video clips randomly sampled from the training split of AudioSet, which is denoted as $\text{Audioset-240K}^\dagger$. The validation and test sets are sampled from the evaluation split of Audioset, which contain 10K and 8K video clips, respectively. Audioset is always used for audio event recognition, but we do not use any annotations other than the unconstrained videos themselves in our task. We only use the YouTube identifiers to download videos from the website, split each entire video into non-overlapping 10-sec. clips, and discard the clips shorter than ten seconds. Compared with other works, we do not even use the annotations of timestamps. The video clips segmented by the timestamps usually have a specific event while the videos in the real-life environment are complicated. Hence, the dataset obtained by our approach is more difficult and realistic. Any unconstrained video can be used for training, and our model can learn more generic and complex representations.

\paragraph{Training data sampling.}
For consistency and fair evaluation, we follow the same settings in the previous work \cite{owens2018audio}. We sample 4.2-sec. clips at the full frame-rate (29.97Hz) from longer 10-sec. video clips in $\text{Audioset-240K}^\dagger$, and the audio streams are shifted by 2.0 to 5.8 seconds in negative samples. The video frames are first uniformly resized to 256 $\times$ 256, randomly cropped into 224 $\times$ 224, and then passed through to the visual encoder. The inputs to the audio encoder are 21 kHz stereo sounds, which do not need to be converted to spectrograms. 

\paragraph{Optimization details.}
We train our models end-to-end on 4 GPUs (16GB P100) for 750,000 iterations with a batch size of 32. We use the Stochastic Gradient Descent (SGD) algorithm with a momentum rate of 0.9 to optimize our model, where the initial learning rate is 0.01 with the weight decayed by $10^{-5}$. To accelerate the training processing, we pre-train our model at a lower video frame-rate of 7.5 Hz and a larger batch size of 64 for the first 100,000 iterations as the warm-up process \cite{he2016deep}, which takes roughly 46 hours. 

\subsection{Results and Discussions}
We evaluate the performance of our approach on the AVS task, which is directly compared with two baseline models \cite{owens2018audio}, including \textit{Multisensory} and \textit{Multisensory*}.
\textit{Multisensory} is the checkpoint provided by \citet{owens2018audio}. This model is pre-trained on the subset of Audioset, which includes approximately 750,000 videos, and we denote this subset as Audioset-750K. 
\textit{$\text{Multisensory}^*$} is the model re-trained on the $\text{Audioset-240K}^\dagger$ (our dataset) to ensure a fair comparison.

\begin{table}[]
\centering
\caption{The experimental results on the AVS task. $\dagger$ denotes the video clips in the dataset are segmented by our methods. * denotes the re-trained model. The notations of \textit{Ours (CMA)}, \textit{Ours (AGA)}, and \textit{Ours (VGA)} denote our co-attention module which consists of both AGA and VGA (i.e., CMA), only AGA, and only VGA, respectively.}
\label{tab:shift-table}

\resizebox{\linewidth}{!}{%
\begin{tabular}{@{}l|c|c|c|c@{}}
\toprule
\multirow{2}{*}{\textbf{Methods}} & \multirow{2}{*}{\textbf{Parameters}} & \multirow{2}{*}{\textbf{Training Set}} & \multicolumn{2}{c}{\textbf{Evaluation (Acc.)}} \\ \cmidrule(l){4-5} 
 &  &  & \multicolumn{1}{l|}{Audioset-750K} & \multicolumn{1}{l}{$\text{Audioset-240K}^\dagger$} \\ \midrule
$\text{Multisensory}$\cite{owens2018audio} & 36M & Audioset-750K & \multicolumn{1}{c|}{59.9\%} & 59.3\% \\
$\text{Multisensory}^*$\cite{owens2018audio} & 36M & $\text{Audioset-240K}^\dagger$ & \multicolumn{1}{c|}{60.2\%} & 59.5\% \\ \midrule
Ours (AGA) & 15M & $\text{Audioset-240K}^\dagger$ & \multicolumn{1}{c|}{61.9\%} & 61.3\% \\
Ours (VGA) & 15M & $\text{Audioset-240K}^\dagger$ & \multicolumn{1}{c|}{62.6\%} & 62.0\% \\
Ours (CMA) & 15M & $\text{Audioset-240K}^\dagger$ & \multicolumn{1}{c|}{\bf{65.3\%}} & \bf{62.6\%} \\ \bottomrule
\end{tabular}%
}
\end{table}

\begin{figure*} [!ht]
\centering
  \includegraphics[width=0.92\textwidth]{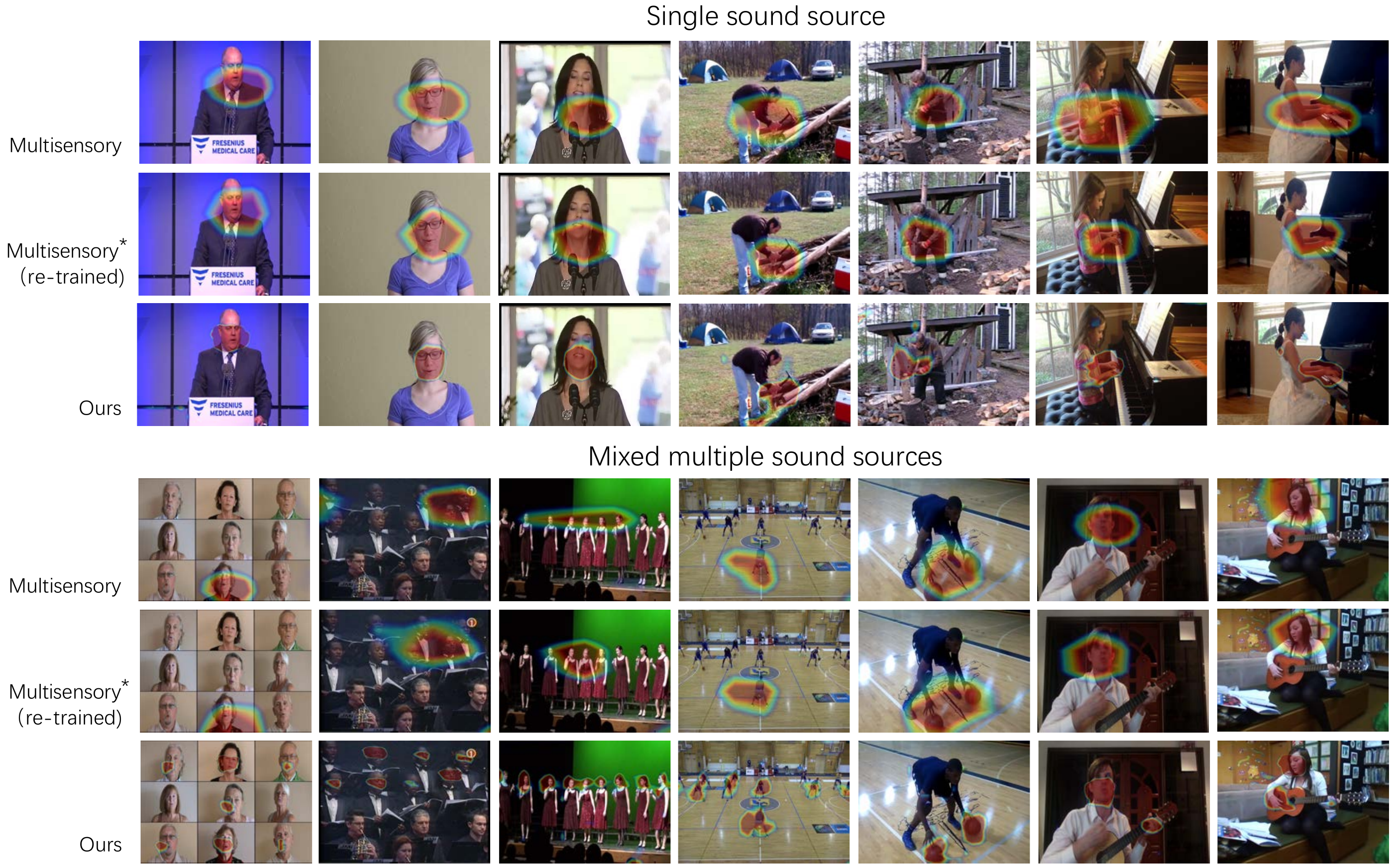}
  \caption{The qualitative results of sound source localization. The first three speech examples are sampled from Audioset and the rest ones are sampled from Kinetics-Sounds (e.g., chopping wood, singing, etc.). The top three rows are the cases that each video contains only one sound source, and the bottom three rows are the cases that each video contains multiple sound sources. More visualized videos can be found at \textit{\textcolor{magenta}{\url{https://youtu.be/UPrZ5Kr-DwA}}}\textcolor{magenta}{.}
  }
  \label{fig:sound_source_compare}
\end{figure*}

The related experimental results are reported in Table \ref{tab:shift-table}, which indicate that our proposed co-attention models achieve superior performance over the baseline models. We observe that our CMA-based model makes obvious improvements over the two ablated models, i.e., AGA-based and VGA-based models. Thus we just discuss the CMA-based model in the rest of this paper.
Our CMA-based model outperforms \textit{$\text{Multisensory}^*$} by 5.1\% and 3.1\% on Audioset-750K and $\text{Audioset-240K}^\dagger$, respectively. This verifies the effectiveness of introducing the cross-modal interactions between audio and visual streams. It should be noted that such performance is achieved with 58.3\% fewer parameters. It is significantly more parametric-efficient than the baseline models. 
Besides, we note that compared to \textit{$\text{Multisensory}$}, the re-trained \textit{$\text{Multisensory}^*$} improves the performance by 0.3\% and 0.2\% on Audioset-750K and $\text{Audioset-240K}^\dagger$, respectively. This demonstrates that the unconstrained videos in our dataset enforce the model to learn more effective and generic representations with less training data.

To study the transferability of our model and further understand which categories the model can achieve good performance, we also evaluate the prediction accuracy of the AVS task on the Kinetics dataset \cite{kay2017kinetics} (without re-training). We find that our CMA-based model still outperforms \textit{$\text{Multisensory}^*$} by a significant margin (60.8\% vs. 57.5\% accuracy). The results of each category are shown in Figure \ref{fig:shift_test}. 
It can be viewed that the most successful predictions are the categories containing human speech (e.g., news anchoring, answering questions), which indicates that our co-attention model is sensitive to lip motions and can be applied to some speech applications (e.g., active speaker detection and speech separation). The worst predictions are some actions that almost make no sound, e.g., slacklining and climbing ladder. In these cases, even human beings can hardly make a correct judgment.

It is worth noting that the AVS task is quite difficult. First, the audio streams may be silent, or the video frames may be unchanged. Second, the unconstrained videos may contain mixed multiple components, making it hard to associate with the real audio-visual pairs and achieve acceptable performance. Moreover, the sound-maker may not even appear on the screen (e.g., the voiceover of photographer, the person narrating the video).
\citet{owens2018audio} gave 30 Amazon Mechanical Turk participants 60 aligned/shifted audio-visual pairs from Audioset, which were 15-sec. in length and the audio tracks were shifted by large, 5 seconds, so that they had more temporal semantic contexts to make predictions. However, the human classification result is only $66.6\% \pm 2.4\%$ accuracy. This suggests that our co-attention model helps to bridge the gap to the performance on human evaluation significantly. 

Audio-visual synchronization has extensive applications in our daily life, e.g., determining the lip-sync errors \cite{chung2016out}, detecting signal processing delays in video camera and microphone. However, as a pretext task in self-supervised learning, the performance on the AVS task is not our final objective. We are also interested in the learned cross-modal representations with the expectation that these representations can carry good semantics or structural meanings and eventually facilitate a variety of downstream tasks. 
In the following sections, we present the qualitative and quantitative evaluation on some practical downstream tasks.

\section{Sound Source Localization}

It is essentially impossible for a neural network to effectively perform the Audio-Visual Synchronization (AVS) task, unless it has first learned to find the discriminative visual regions which make the sound. Here, we show how our co-attention model associates the visual regions with the sound components. To this end, we visualize the results by using the Class Activation Map (CAM) \cite{zhou2016learning}, which exploits Global Average Pooling (GAP) \cite{lin2014network} to build effective localizable representations that recognize the visual regions.

The qualitative results are shown in Figure \ref{fig:sound_source_compare}. We compare the model that performs best on the pretext task with the baseline models, i.e., \textit{$\text{Multisensory}$} and \textit{$\text{Multisensory}^*$}. As can be seen, the models can recognize the objects that make the sounds or whose motions are highly correlated to the sounds (e.g., lip motion), which are used for detecting the misalignments between audio and visual streams. The top three rows show the examples with only one sound source in each video. From the results in Figure \ref{fig:shift_test}, we find that the most successful predictions of the AVS task are the categories involving human speech. Hence, we perform sound source localization on the held-out speech subset of Audioset and display the results of three randomly sampled speech instances in Figure \ref{fig:sound_source_compare}. It can be viewed that both multisensory models and our model are sensitive to face and mouth movements. Specifically, the difference between \textit{$\text{Multisensory}$} and \textit{$\text{Multisensory}^*$} is not too much. Whereas the co-attention module in our model can benefit the fine-grained representations of audio and visual streams, leading to localizing the sound sources more precisely and more concentratedly. 

To further demonstrate the generalizability of our model to other categories of videos, we apply the models to the Kinetics-Sounds dataset \cite{arandjelovic2017look} and show some examples in Figure \ref{fig:sound_source_compare}, including \textit{chopping woods, playing organ, singing, dribbling basketball and playing guitar}. We can see that the models can still localize the action regions.

The bottom three rows in Figure \ref{fig:sound_source_compare} are more complex and challenging cases that each video contains multiple sound sources. Compared with the baseline models, our model can almost localize each sound source in various categories of videos. For example, as shown in the fourth column, one coach and some athletes are dribbling basketball in the gym. Even though they have similar actions that make the same thuds of bouncing balls, our co-attention model still can localize each athlete and the coach precisely, whereas the baseline models just localize the visual region of the coach and ignore the actions of athletes. These results indicate that the multi-head attention mechanism facilitates the network to capture various information, attend to discriminative visual regions correlated to the sound sources, and localize the regions in the scenes. 

\section{Action Recognition}
To evaluate the effectiveness of cross-modal representations that emerge from the pretext task quantitatively, we fine-tune our model on two standard action recognition datasets of UCF101 \cite{soomro2012ucf101} and HMDB51 \cite{Kuhne2011hmdb}. UCF101 consists of 13K videos from 101 human action classes, and HMDB51 contains about 7K clips from 51 different human motion classes. UCF101 and HMDB51 have three different official training/testing split lists, and the mean accuracy over the three splits are computed during our experiments. We compare our model with the fully-supervised 3D CNN methods and other self-supervised methods, and discuss the results in this section.

For action recognition, the task requires the models to classify the action labels of given videos. We concatenate our two final output audio and visual features of co-attention modules and add two fully-connected layers on the top of the pre-trained model for classification. The dimension of the last fully-connected layer is the number of labels in the dataset. We fine-tune the entire model with cross-entropy loss, and the weights are initialized by the model pre-trained on the AVS task.

\begin{table}[]
\caption{The action recognition accuracy on UCF101 \cite{soomro2012ucf101} and HMDB51 \cite{Kuhne2011hmdb}. We compare our model against the fully-supervised 3D CNN methods and other self-supervised audio-visual methods.}
\label{tab:action-table}
\resizebox{\linewidth}{!}{%
\begin{tabular}{@{}lllll@{}}
\toprule
\multicolumn{1}{c|}{\multirow{2}{*}{Methods}} & \multicolumn{2}{c|}{Pre-training} & \multicolumn{2}{c}{Evaluation} \\ \cmidrule(l){2-5} 
\multicolumn{1}{c|}{} & \multicolumn{1}{l|}{Dataset} & \multicolumn{1}{l|}{Size} & \multicolumn{1}{l|}{UCF101} & HMDB51 \\ \midrule
\multicolumn{1}{l|}{I3D-RGB \cite{carreira2017quo}} & \multicolumn{1}{l|}{None} & \multicolumn{1}{l|}{0K} & \multicolumn{1}{l|}{57.1\%} & 40.0\% \\
\multicolumn{1}{l|}{I3D-RGB \cite{carreira2017quo}} & \multicolumn{1}{l|}{Kinetics} & \multicolumn{1}{l|}{240K} & \multicolumn{1}{l|}{95.1\%} & 74.3\% \\
\multicolumn{1}{l|}{I3D-RGB \cite{carreira2017quo}} & \multicolumn{1}{l|}{Kinetics + Imagenet} & \multicolumn{1}{l|}{-} & \multicolumn{1}{l|}{95.6\%} & 74.8\% \\ \midrule
\multicolumn{1}{l|}{$\text{Multisensory}^*$ \cite{owens2018audio}} & \multicolumn{1}{l|}{$\text{Audioset-240K}^\dagger$} & \multicolumn{1}{l|}{240K} & \multicolumn{1}{l|}{82.9\%} & 54.8\% \\
\multicolumn{1}{l|}{XDC \cite{alwassel2019self}} & \multicolumn{1}{l|}{Kinetics} & \multicolumn{1}{l|}{240K} & \multicolumn{1}{l|}{84.2\%} & 47.1\% \\
\multicolumn{1}{l|}{AVTS \cite{korbar2018cooperative}} & \multicolumn{1}{l|}{Kinetics} & \multicolumn{1}{l|}{240K} & \multicolumn{1}{l|}{85.8\%} & 56.9\% \\ \midrule
\multicolumn{1}{l|}{Ours (scratch)} & \multicolumn{1}{l|}{None} & \multicolumn{1}{l|}{0K} & \multicolumn{1}{l|}{70.5\%} & 44.8\% \\
\multicolumn{1}{l|}{Ours (vision only)} & \multicolumn{1}{l|}{$\text{Audioset-240K}^\dagger$} & \multicolumn{1}{l|}{240K} & \multicolumn{1}{l|}{83.3\%} & 51.9\% \\
\multicolumn{1}{l|}{Ours (full)} & \multicolumn{1}{l|}{$\text{Audioset-240K}^\dagger$} & \multicolumn{1}{l|}{240K} & \multicolumn{1}{l|}{\textbf{87.8\%}} & \textbf{58.2\%} \\ \midrule
\multicolumn{5}{l}{\textit{Using larger datasets}} \\ \midrule
\multicolumn{1}{l|}{L3-Net \cite{arandjelovic2017look}} & \multicolumn{1}{l|}{Audioset} & \multicolumn{1}{l|}{2M} & \multicolumn{1}{l|}{72.3\%} & 40.2\% \\ \midrule
\multicolumn{1}{l|}{Multisensory \cite{owens2018audio}} & \multicolumn{1}{l|}{Audioset-750K} & \multicolumn{1}{l|}{0.75M} & \multicolumn{1}{l|}{82.1\%} & 54.0\% \\ \midrule
\multicolumn{1}{l|}{AVTS \cite{korbar2018cooperative}} & \multicolumn{1}{l|}{Audioset} & \multicolumn{1}{l|}{2M} & \multicolumn{1}{l|}{89.0\%} & 61.6\% \\ \midrule
\multicolumn{1}{l|}{XDC \cite{alwassel2019self}} & \multicolumn{1}{l|}{Audioset} & \multicolumn{1}{l|}{2M} & \multicolumn{1}{l|}{91.2\%} & 61.0\% \\ \midrule
\multicolumn{1}{l|}{XDC \cite{alwassel2019self}} & \multicolumn{1}{l|}{IG65M} & \multicolumn{1}{l|}{65M} & \multicolumn{1}{l|}{94.2\%} & 67.4\% \\ \bottomrule
\end{tabular}
}
\end{table}

\subsection{Implementation Details}
\paragraph{Data sampling.}

During the fine-tuning procedure, the inputs to the visual encoder are video frames, which are resized to $256 \times 256$, and then center-cropped into $224 \times 224$. At the training stage, we choose the starting frame randomly and then sample 2.56-sec. video clips. At inference, we follow \cite{simonyan2014two} and sample 25 overlapping subclips with equal temporal spacing between them. We compute the outputs for each video subclip and average their outputs to make a video-level prediction.

\begin{figure*} [!ht]
\centering
  \includegraphics[width=0.85\textwidth]{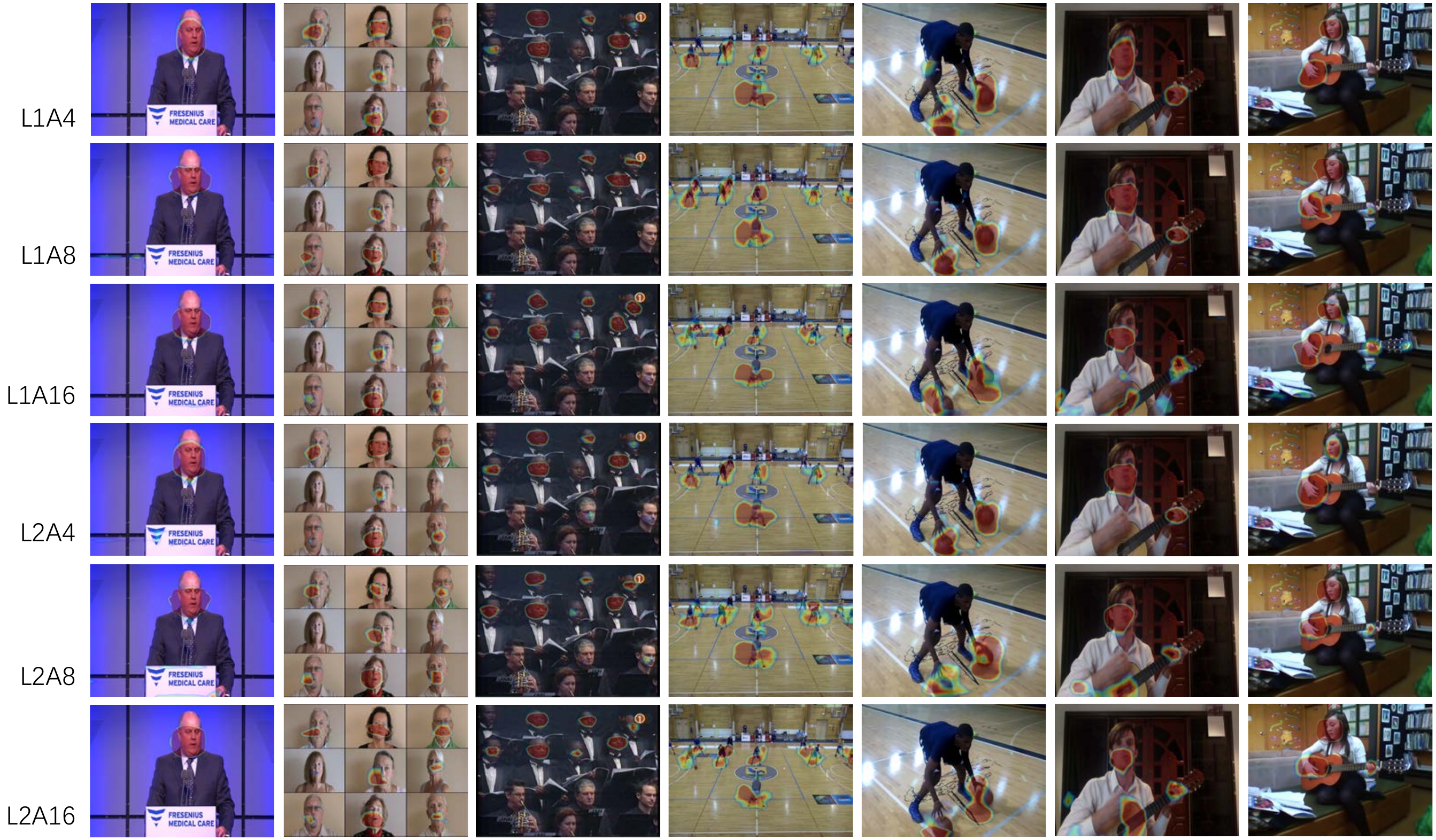}
  \caption{The ablation study results on the task of sound source localization. The first example contains only one sound source and the rest examples contain mixed multiple sources.}
  \label{fig:ablation_study}
\end{figure*}

\paragraph{Optimization details.}
We fine-tune our models on single 16G P100 for about 30,000 iterations. The batch size is set as 16. We use Adam to optimize our models with the initial learning rate of 3e-5, and halve the rate every 10,000 iterations. Dropout is imposed with a probability of 0.5. Data augmentation strategies, i.e., random cropping and shifting audio clips, are applied to reduce overfitting. 

\subsection{Results and Discussions}
The comparison results are reported in Table \ref{tab:action-table}. We observe that our co-attention model provides a remarkable boost against other self-supervised methods when pre-trained on the same size of dataset. Our model improves by 2.0\% on UCF101 and 1.3\% on HMDB51 compared with AVTS \cite{korbar2018cooperative} pre-trained on Kinetics (240K examples). It is worth noting that the visual subnetwork adopted in AVTS is based on the mixed-convolution (MCx) family of architectures \cite{tran2018closer}, which is designed for action recognition and has more parameters than our model. We also observe that the accuracy grows monotonically as the size of pre-training dataset increases, which suggests that our co-attention model may reduce the remaining gap to the fully-supervised method by using more pre-training data.

To evaluate the effectiveness of the self-supervised pre-training mechanism, we also train the neural network of the same cross-modal architecture from scratch. We can see that our full model improves by 17.3\% on UCF101 and 13.4\% on HMDB51 when pre-trained on the pretext task, which verifies our self-supervised pre-training is beneficial to the task of action recognition.

We further explore whether the pre-trained visual subnetwork (vision-only) can work alone. For that purpose, we set the activation of audio streams to zero and change the co-attention to self-attention, that is, modify the queries $Q_{v}$ in the video streams to the intermediate video features $F_{v}$. We observe that there is a 4.5\% fall on UCF101 and a 6.3\% fall on HMDB51, respectively, which proves the audio subnetwork is important for action recognition and the visual subnetwork can work independently.

\begin{table}[!htp]
\centering
\caption{The ablation study results on the tasks of Audio-Visual Synchronization (AVS) and action recognition. \#L denotes the depth of co-attention module, and \#A denotes the number of attention heads.}
\label{tab:ablation}

\resizebox{0.7\linewidth}{!}{%
\begin{tabular}{@{}ll|ccc@{}}
\toprule
\multicolumn{2}{l|}{\textbf{Hyperparams}} & \multicolumn{3}{c}{\textbf{Evaluation (Acc.)}} \\ \midrule
\#L & \#A & \multicolumn{1}{c|}{AVS} & UCF101 & HMDB51 \\ \midrule
1 & 4 & \multicolumn{1}{c|}{64.2\%} & 85.6\% & 57.0\% \\
1 & 8 & \multicolumn{1}{c|}{\textbf{65.3\%}} & \textbf{87.8\%} & \textbf{58.2\%} \\
1 & 16 & \multicolumn{1}{c|}{64.6\%} & 87.4\% & 57.9\% \\
2 & 4 & \multicolumn{1}{c|}{64.1\%} & 84.5\% & 55.1\% \\
2 & 8 & \multicolumn{1}{c|}{64.8\%} & 86.8\% & 56.9\% \\
2 & 16 & \multicolumn{1}{c|}{64.3\%} & 85.9\% & 56.6\% \\ \bottomrule
\end{tabular}%
}

\end{table}

\section{Ablation Study}
\label{ablation_study}

In this work, we denote the hidden size of transformer as \emph{H}, the number of attention heads as \emph{A}, and the depth of co-attention module as \emph{L}. In this section, we explore the effects on six model sizes, i.e., \emph{L}=[1,2], \emph{H}=512, \emph{A}=[4,8,16].

Table \ref{tab:ablation} shows the results of ablation study on the Audio-Visual Synchronization (AVS) and action recognition tasks with respect to different depths and heads of the model. We find that both of these two tasks benefit from shallower models, and increasing the number of heads properly can also improve the performance. Figure \ref{fig:ablation_study} presents the effects on localizing the sound sources. We observe that all models perform similarly when the video contains only one sound source. For better comparison, we show more examples that contain mixed multiple sounds. It can be seen that the greater depth of co-attention module allows the model to localize the sound source more concentratedly. We also observe that the model can localize more sound sources with the increase of head number. For example, in the second column, there are nine older people singing on the screen. As the number of heads increases, the model can identify the older people in the bottom left corner and middle right. 
This result verifies that the multi-head mechanism is helpful to localize multiple visual regions correlated to the sounds.

\section{Conclusion and Future Work}
In this paper, we propose a self-supervised framework with co-attention mechanism to learn generic cross-modal representations by solving the pretext task of Audio-Visual Synchronization (AVS). Our co-attention model introduces the information interactions between audio and visual streams, which can achieve state-of-the-art performance on the AVS task. We also demonstrate that it successfully learns cross-modal semantic information that can be utilized for a variety of downstream tasks, such as sound source localization and action recognition. Note that we train our networks with a subset of Audioset for efficiency purposes, we will exploit more pre-training data to achieve continual improvement, and apply our model to other practical audio-visual tasks in future work.

\begin{acks}
This work was partially supported by National Natural Science Foundation
of China (No. 61976057, No. 61572140), Military Key Research Foundation Project (No. AWS15J005), Shanghai Municipal Science and Technology Major Project (2018SHZDZX01) and ZJLab.
\end{acks}

\bibliographystyle{ACM-Reference-Format}
\bibliography{sample-base}


\end{document}